# A generalized deep learning model for multi-disease Chest X-Ray diagnostics

Nabit A. Bajwa[1,2][†], Kedar Bajwa[1][§], Atif Rana[1][‡], Muhammad Faique Shakeel[1][¶], Kashif Haqqi[1,2][||] &

Suleiman Ali Khan[3][*]

[1]*Shifa International Hospital, Islamabad, Pakistan, 44000*

[2]*George Mason University, Fairfax, VA, United States of America*

[3]*Institute for Molecular Medicine Finland FIMM, University of Helsinki, Helsinki, FI-00014, Finland*

**We investigate the generalizability of deep convolutional neural network (CNN) on the task of disease classification from chest x-rays collected over multiple sites. We systematically train the model using datasets from three independent sites with different patient populations: National Institute of Health (NIH), Stanford University Medical Centre (CheXpert), and Shifa International Hospital (SIH). We formulate a sequential training approach and demonstrate that the model produces generalized prediction performance using held out test sets from the three sites. Our model generalizes better when trained on multiple datasets, with the CheXpert-Shifa-NET model performing significantly better (p-values < 0.05) than the models trained on individual datasets for 3 out of the 4 distinct disease classes. The code for training the model will be made available open source at: www.github.com/link-to-code at the time of publication.**

[†]E-mail: nbajwa4@gmu.edu
[§]E-mail: kedar.bajwa@gmail.com
[‡]E-mail: atif.rana@shifa.com.pk
[¶]E-mail: muhammadfaiques@gmail.com
[||]E-mail: dockash@gmail.com
[*]E-mail: suleiman.khan@helsinki.fi

**Declaration of interest:** none

# 1  Introduction

Deep learning image classification models are being widely researched and adopted in a variety of use cases in the healthcare domain. Whether it is for screening patients with blinding retinal diseases[1], classifying skin cancer using deep learning[2], or detecting pneumonia from chest x-rays[3], deep learning has outperformed traditional computer vision techniques in nearly all accuracy benchmarks in image classification tasks[4,5].

Globally, Chest X-rays are projected to account for nearly 27% of all diagnostic medical x-ray examinations[6], and in many cases, is the first examination which is prescribed to either diagnose or monitor a wide variety of lung or chest diseases[7]. These diseases range from pneumonia, to lung cancer, heart failure or pneumothorax. Chest x-rays can also be used to investigate a case of suspected tuberculosis, or interstitial lung diseases. Hence, it holds a valuable role in the diagnosis of patients suspected of multiple disorders and coupled with the near universal availability of the test, it is one of the major diagnostic tools available in current medicine practice.

There are several factors which contribute to deep learning being a good choice for the analysis of chest x-rays, and for the development of Computer Aided Diagnostic (CAD) tools which benefit from it. In hospitals all over the world, hundreds of millions of chest radiological examinations have been conducted and thousands continue to be conducted daily[6]. With such a huge number of chest x-ray images existing in hospital databases worldwide, it provides a natural use case for deep learning techniques. While in theory, the combination of deep learning with chest x-rays should produce results that are on par with the use of deep learning in other image classification, recognition and segmentation tasks, there are a few limitations which hinder this progress. Deep learning techniques, while providing unmatchable accuracies in image classification tasks require a large amount of labeled images to achieve this. While images of cats, dogs, cars and other normal everyday objects are very easy to obtain and label, challenges exist when it comes to obtaining and labeling medical data. Medical data is protected by strong patient confidentiality compliance requirements, and moreover medical images rarely exist with single one-word labels attached to them. Ideally, medical image reports should be labeled with ICD codes[8], such data is not widely profiled. Generally, medical images are coupled with a corresponding radiologist report which details the background, observations and findings of the patient in that particular study.  These radiology reports often vary in structure and technique across radiologists and most certainly across sites. Thus, to make these radiological images usable at large scale for training machine learning algorithms, we require radiologists to manually label these images, or a methodology to mine disease labels from these radiology reports. Since, manual annotation of thousands of images is an extremely laborious and resource intensive task, researchers primarily rely on disease label mining from the radiology report[9–11]. This label mining may introduce noise in the labeling process. However, in the case of medical imagery, existing work suggests that despite the noise introduced by automatic labeling, deep learning models can learn important features for diagnostic purposes much better than classical computer vision techniques[3].

In this work, we study the generalizability of deep convolutional networks for chest x-ray classification. Our main contributions in this work is an analysis of the generalizability of deep learning models when trained on chest x-rays from multiple sites and over multiple disease classes. We robustly evaluate generalization performance across multiple disease classes and multiple datasets from different patient populations.

Our motivation for this work is to gain an understanding in how weakly labeled medical image data can be used to effectively train deep learning models, and to study the effect datasets from external sites have on the accuracy and generalizability of these models. Given that there is a tremendous amount of medical imagery data stored in hospital sites worldwide, we take a step towards building a cost-effective, accurate and generalizable methodology to train chest x-ray classifier. Our proposed methodology is accurate, easy to use and extend to additional sites sequentially.

## 2 Related Works

Recently, disease classification from chest x-rays has been a focus area for researchers, with multiple studies showing deep learning techniques being used to detect diseases from chest x-rays[12,13]. As a seminal work in the domain, NIH Chest X-Ray8[13] was a large open source dataset of frontal chest x-rays, with more than a hundred thousand chest x-rays and labels provided for 14 different disease classes. The dataset accelerated the interest in deep learning algorithms for disease detection in chest x-rays, and inspired several works. The authors[13] also introduced a deep convolutional network based on ResNET50 for x-ray disease classification. These works ranged from developing different architectures for single and multiple disease detection[14], to localizing[15] and even partially writing the radiology reports using recurrent neural networks[16].

CheXNet[3] and CheXNeXt[17] claim radiologist-level disease detection on multiple disease classes from chest x-rays. Both of these deep learning models were trained on the NIH dataset, and a comparison was drawn between the two models and board-certified radiologists on a held-out test set from the NIH dataset over multiple disease classes. Following the release of these two models, researchers from the same group released an open source dataset of chest x-rays, CheXpert[18], which, coupled with the MIMIC-CXR[19] dataset is the largest dataset of chest radiographs to date. While the MIMIC-CXR can be obtained only after a lengthy verification and approval process, the CheXpert dataset is available much more easily. CheXpert also provides a radiologist annotated validation set. It contains labels from 14 different disease classes. Our work builds on this massive release of chest radiographs and their corresponding labels; and investigates an important claim: that CNN's generalize to real world data if given enough images of that class[20].

Recently authors in [21] used chest x-ray images from multiple sites to investigate the generalizability of Convolutional Neural Networks (CNN's) for the task of pneumonia detection. This is done by training multiple deep learning models to detect pneumonia using individual and pooled datasets from multiple sites, and compares the internal and external performance for these trained models. The internal performance is defined as the performance of the deep learning model on the test data from the hospital it was trained on, while the external performance is defined as the performance of the model on datasets from hospital sites independent to the one it was trained on. Their work claims that CNN's do not necessarily generalize to unseen test data from external test sites in the case of pneumonia detection. While their work is similar to ours, we differ in two key aspects, that is, our method generalizes to multiple disease classes, and we use to sequential training methodology instead of pooling, specifically we propose a methodology to train and validate models in multiple stages using single datasets to evaluate their generalization ability. Moreover, to investigate generalization performance of automated chest x-ray classification models, authors in [22,23] explored various CNN models. Specifically, they focused on quantifying the generalization ability of CNN models and what the limitations of the CNN models are, by conducting multiple experiments across multiple open source datasets on multiple disease classes. Our work confirms the generalizability of deep learning models for Chest X-Ray datasets analogous to[22], however, we also demonstrate the generalizability on a large Chest X-Ray dataset which is from an entirely different patient population coming from Asian background (SIH), as well as validation of the findings by medical experts (radiologist).

## 3  Methods and Datasets

We investigate the effect of using X-Ray datasets from multiple sites on the generalizability of deep learning models. We conduct experiments over multiple disease classes to assess the robustness of the approach. Specifically, a model trained on the NIH Chest X-Ray8 dataset[13] was used as a baseline. We generalize and train the baseline NIH model with two large independent datasets coming from distinct sites and populations. The datasets have been recently profiled for patient diagnosis and consist of chest X-rays and labels based on corresponding radiologist reports. The two datasets are the CheXpert dataset by Stanford, USA[18] and an internal Chest X-Ray dataset by Shifa International Hospital (SIH), Pakistan. We train three models, namely CheXpert-NET, Shifa-NET and CheXpert-Shifa-NET. The model CheXpert-NET was trained on the CheXpert dataset, Shifa-NET was trained on the SIH internal dataset, and CheXpert-Shifa-NET was trained on on both the CheXpert dataset and the SIH internal dataset, sequentially. The models were evaluated using the Area Under the Receiver Operating Curve (AUROC)[24]. We use independent test sets from each of the sites, NIH, CheXpert and Shifa, to assess the generalizability of the model.

## 3.1 Datasets

In this section, we present the three datasets used in this study. These datasets are collected from independent medical sites, two in the US and one in Pakistan. Data is collected for four disease classes: Atelectasis, Cardiomegaly, Pleural Effusion and Pneumonia. In addition to the four classes, we use a No Finding class, which is a normal chest x-ray and denotes the absence of the diseased classes. All three training datasets are labeled with an automated labeler. The test sets of NIH and Shifa datasets are automatically labeled, while the testing split for the CheXpert dataset provides a strong radiologist annotated ground truth.

### 3.1.1 Shifa International Hospital (SIH) Chest X-ray Dataset

This is an internal dataset consisting of 80,035 frontal chest x-ray images from 43,839 patients in Shifa International Hospitals (SIH), Islamabad, Pakistan. The images collected are from the period of 1st Jan 2014 till 31st December 2018. The images were extracted based on keywords found in the corresponding radiologist reports, and were approved for use in this study by the institutional review board (IRB).

We use an automatic labeler based on disease keywords obtained from multiple domain expert radiologists and a manual review of reports. These keywords are as listed in Table 1.

| Classes | Keywords |
| --- | --- |
| **No Finding** | 'normal study', 'unremarkable study', 'no active lung lesion', 'no active lung disease', 'no acute cardiopulmonary lesion' |
| **Atelectasis** | 'atelectasis', 'atelectatic bands' |
| **Cardiomegaly** | 'cardiomegaly', 'enlarged cardiac size' |
| **Pleural Effusion** | 'bilateral pleural effusion', 'right sided pleural effusion', 'left sided pleural effusion |
| **Pneumonia** | 'air space opacification', 'air space consolidation', 'opacification', 'zone infiltrates' |

*Table 1: Keywords for Shifa International Hospital reports classification*

Given the unstructured nature of radiology reports and the variation in writing style for different radiologists, we assume our images to be weakly labeled at best. We do not employ any strategy for negation detection or uncertainty detection in our labeling methodology. The labeling technique used classifies each image based on the presence of the given key-words in each radiology report. If the given keyword is present in the radiology report, the image is labeled as positive for that class. Hence, our dataset is a multi-labeled dataset which contains a label for each of the four classes; positive or negative. Table 2 displays a sample of the reports and corresponding labels attributed to it by our automatic labeler.

| Report Text | Labels |
| --- | --- |
| In place NG tube and right-sided CVP line are again noted. Minimal left **pleural effusion** with underlying lung consolidation/**atelectasis** is again noted. Infiltrates are again noted in bilateral upper and right mid zones. Minimal atelectatic changes in the right lung base are seen. Rest of the findings are unchanged. " | Atelectasis, Pleural Effusion |
| There are persistent inhomogenous areas of **airspace opacification** in right lung. Mild interval increase in the inhomogenous areas of **airspace opacification** in left lung is noted. There is interval development of mild inhomogenous haziness in bilateral lower zone obscuring both CP angles and diaphragmatic outlines suggesting **pleural effusions** with underlying **atelectasis** on left. | Atelectasis, Pleural Effusion, Pneumonia |
| Portable technique and AP position are limiting details. There is moderate **right pleural effusion** with underlying consolidation or **atelectasis**. Mild left **pleural effusion** is also noted. There is accentuation of bronchovascular markings in both lungs. No pneumothorax is present. Cardiomediastinal magnification may be due to AP projection however requires follow-up with proper positioning to rule out **cardiomegaly**. Degenerative changes are noted in the bones. Soft tissues are unremarkable. | Atelectasis, Pleural Effusion, Cardiomegaly |
| There is re-demonstration of prominent interstitial markings. No collapse or consolidation is noted. There is no pneumothorax. There is persistent blunting of left CP angle likely due to pleural thickening or **pleural effusion**. Right CP angle is normal. **Cardiomegaly**. cardiac pacemaker with its intact leads are again noted. Rest of the findings are unchanged. " | Cardiomegaly, Pleural Effusion |
| No definite evidence of any consolidation, collapse or pneumothorax is seen. Both hila and cardiomediastinal contours appear normal. Both CP angles are acute. Both hemidiaphragm appear normal. No evidence of any bony lesion seen. CONCLUSION: **Normal study**. | No Finding |

*Table 2: Sample SIH radiology reports and their corresponding labels*

We use 85/10/5 splits for our training, validation, and testing sets. The data splits drawn are completely random. The total number of images and per class prevalence of images in each of the data splits are given in Table 3, and are fairly balanced.

|  | Train | Validation | Test | Total |
|---|---|---|---|---|
| **Atelectasis** | 14,830 | 1,730 | 874 | 17,434 |
| **Cardiomegaly** | 10,015 | 1,208 | 592 | 11,815 |
| **Pleural Effusion** | 22,350 | 2,627 | 1,339 | 26,316 |
| **Pneumonia** | 14,653 | 1,758 | 885 | 17,296 |
| **No Finding** | 7,948 | 876 | 456 | 9,280 |

*Table 3: Per class prevalence for disease positive cases in training, testing and validation splits in SIH dataset*

### 3.1.2 National Institute of Health (NIH) Chest X-ray8 Dataset

The NIH Chest X-Ray8[2] dataset consists of 112,120 X-ray images with disease labels from 30,805 unique patients. It provides labeled data for 14 different diseases. The images have been labeled using an automated labeling technique.

We use a model pretrained on NIH dataset as a baseline model. This baseline model is an implementation[25] of the DenseNet-121[26] model trained on the NIH dataset for 14 classes. Given the wider generalizability of DenseNet-121 model, we preferred this architecture over NIH's original ResNet architecture. Hence, we report the data splitting strategy used by the corresponding implementation of the DenseNet-121 model [25]in our work. Table 4 denotes the per class prevalence in the data splits.

|  | Train | Validation | Test | Total |
|---|---|---|---|---|
| **Atelectasis** | 10,768 | 652 | 139 | 11,559 |
| **Cardiomegaly** | 2,564 | 155 | 57 | 2,776 |
| **Pleural Effusion** | 12,381 | 737 | 199 | 13,317 |
| **Pneumonia** | 1,328 | 79 | 24 | 1,431 |
| **No Finding** | 56,203 | 3,380 | 778 | 60,361 |

*Table 4: Per class prevalence for disease positive cases in training, validation and testing splits in NIH dataset*

Moreover, we combine the validation and test splits for testing our models in our experimental setup, to get a higher number of test samples.

### 3.1.3 CheXpert Stanford Dataset

CheXpert[18] is a large public dataset released by Stanford University for chest radiograph interpretation. It consists of 224,316 chest radiographs and corresponding disease labels from a total of 65,240 patients, with the images being collected from chest x-ray examinations performed between October 2002 and July 2017 in both inpatient and outpatient centers from Stanford Hospital. The images are released with labels for 14 classes, and three labels per class i.e. positive, negative and uncertain.

For our work, we use only images that are labeled positive, negative and uncertain for 4 radiological observations i.e Atelectasis, Cardiomegaly, Pleural Effusion, and Pneumonia. Moreover, we use the U-Ones strategy from CheXpert work[18] to map uncertain labels to positive. This U-Ones labeling strategy was selected for our experimental setup as it matches closely to the labeling strategy on the other datasets. Specifically, similar to U-one strategy, our internal SIH Chest X-Ray dataset does not use negation detection or uncertainty detection.

Furthermore, as the authors of the CheXpert[18] have not released a public test set, we use the radiologist labeled validation set of the CheXpert dataset as a testing set for our experimentation. We further manually split their training set 95/5 to create our training and validation sets. Therefore, in this study, the CheXpert test set will refer to the human annotated validation set provided by CheXpert. The per-class prevalence in the new training and validation split after applying the U-Ones labeling strategy, along with the test set class prevalence are detailed in Table 5.

|  | **Train** | **Validation** | **Test** | **Total** |
| --- | --- | --- | --- | --- |
| **Atelectasis** | 63,810 | 3,305 | 80 | 11,559 |
| **Cardiomegaly** | 33,283 | 1,804 | 68 | 2,776 |
| **Pleural Effusion** | 92,903 | 4,912 | 67 | 13,317 |
| **Pneumonia** | 23,523 | 1,286 | 8 | 1,431 |
| **No Finding** | 21,235 | 1,146 | 38 | 60,361 |

*Table 5: Per class prevalence for disease positive cases in training, validation and testing splits in CheXpert dataset*

### 3.2 Methods

Given multiple datasets $X^{(1)}$, $X^{(2)}$, and $X^{(3)}$ each corresponding to X-ray images of multiple patients and the corresponding disease labels $Y^{(1)}$, $Y^{(2)}$, and $Y^{(3)}$, we formulate a methodology to train a generalized DenseNet-121[26] model for multi-label disease prediction. DenseNets are a class of convolutional neural networks in which the key idea is that feature maps from the previous layers are concatenated into the inputs of the future layers. While traditional convolutional neural networks have N layers with N connections (one between each layer and the next one), DenseNets have N(N+1)/2 connections. Despite having more connections between layers, DenseNets have fewer parameters, as this structure eliminates the need to relearn redundant feature maps. Moreover, these networks do not suffer from the vanishing gradient problem and have better feature propagation[26].

For each training instance, we optimize the binary cross entropy loss given as:

$$L(y, \hat{y}) = -\frac{1}{5}\sum_{i=0}^{5}(y * \log \hat{y}_i + (1-y) * \log(1 - \hat{y}_i))$$

The trained model is evaluated using the per class AUROC metric on the validation set. The model is trained end to end using the hyperparameters as defined in the experimental setup. We train multiple models in two stages, as detailed in Figure 1. Using a pretrained DenseNet-121 model as a baseline, which we use as an initialization in the first stage, we train two models individually on the CheXpert dataset (namely CheXpert-NET) and the SIH dataset (Shifa-NET). In the second stage, CheXpert-NET is used as an initialization, and trained on the SIH dataset to create a final generalized model (CheXpert-Shifa-NET). To assess the performances of the models, AUC scores are computed on the test set for all three models over the four disease classes.

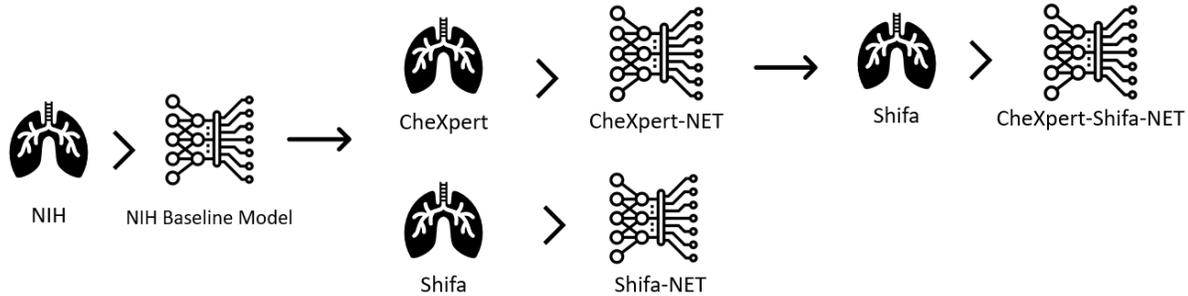

Figure 1: Multi-stage model training flow

### 3.3 Experimental Setup

In this section, we describe our experimental setup. Our methodology for training, evaluating and testing the models is summarized in Figure 2. For each dataset, we start by defining train, test and validation splits. Our models are trained end-to-end updating all model layers for 30 epochs. The models are compiled using binary cross-entropy loss with the Adam optimizer[27] using the default parameters. The initial learning rate was set at 1e-3, with a decay of factor of 10 if there is no improvement in validation loss for 5 continuous epochs. The batch size was set to 16 for all model trainings. During training, data transformations and augmentations were applied on the fly using the Keras Image Data Generator[28] class. The images were resized to a size of 320 by 320, randomly horizontally flipped and randomly rotated by a maximum range of 20 degrees. Pixels were also randomly shifted horizontally and vertically by a factor of 0.20. This level of data augmentation was selected to be part of our experimental setup after careful experimentation, as it helped the model converge faster and achieved lower validation loss scores. The model was checkpointed and validation AUC scores were computed as well at the end of each epoch. All training and testing was done on a local machine which was configured with a NVIDIA 1070 Ti GPU, 16 GB of RAM and an Intel Xeon processor.

The best epoch for these fully trained models is selected for testing based on the lowest validation loss and highest validation mean AUC scores on the 4 categories, details of which can be found in the supplementary materials section of this paper. These fully trained models are evaluated on all three test sets. The code for training the model will be made available open source at: www.github.com/link-to-code.

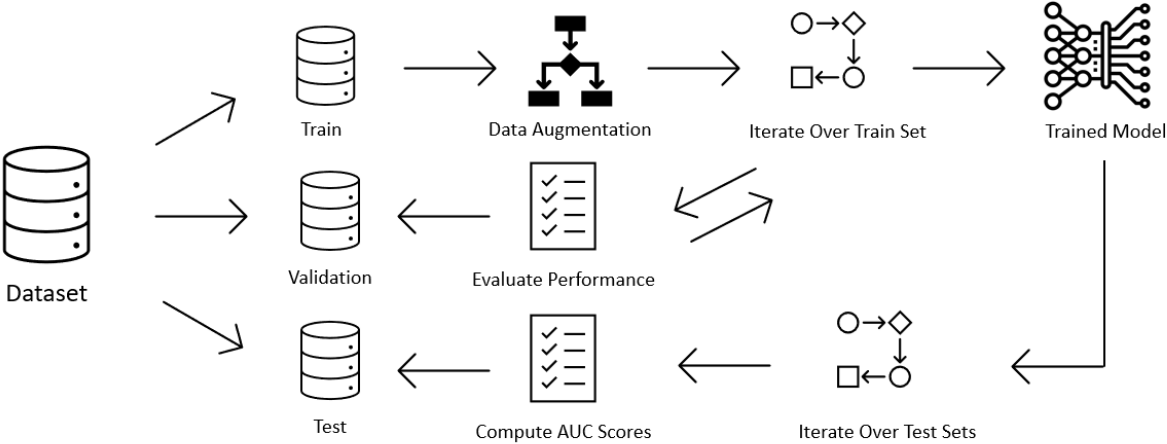

*Figure 2: Experimental methodology*

## 4 Results

We first evaluate our models performance on generalizability task. Specifically, we trained 3 models, CheXpert-NET, Shifa-NET and CheXpert-Shifa-NET, and tested the performance of each of the models on test sets obtained from the three different sources.

We first discuss the overall performance, calculated as the average AUC over all four diseases. This is followed by a detailed performance evaluation and discussion of the models on each disease class. Table 6 demonstrates the average AUC scores of the models on the three datasets. We note that the CheXpert-Shifa-NET model shows a significant improvement over the baseline model than the Shifa test set and the CheXpert test. This improvement is greater in magnitude then CheXpert-NET and Shifa-NET on both test sets, validating that the model learns generalizable features useful for disease prediction. On the NIH test set, the improvement is second to CheXpert-NET model. On average, the CheXpert-Shifa-NET model outperforms all the other three models, attesting that the generalized model is most suitable for disease prediction on arbitrary test images.

|  | Shifa Test Set | CheXpert Test Set | NIH Test | Average |
|---|---|---|---|---|
| **NIH Baseline** | 0.777 | 0.815 | 0.792 | 0.795 |
| **CheXpert-NET** | 0.790 | 0.824 | **0.821** | 0.812 |
| **Shifa-NET** | 0.840 | 0.825 | 0.805 | 0.823 |
| **CheXpert-Shifa-NET** | **0.842** | **0.834** | 0.806 | **0.828** |

*Table 6: Average testing set results of the 4 models on the 4 disease classes*

Table 7 lists detailed prediction performance (AUC scores) of all four models across the four disease classes and the three testing data sets. For the average performance on each disease class, the standard deviation is also computed. The results demonstrate that CheXpert-Shifa-NET is the best individual model and outperforms the other models for most disease-dataset combinations. For three out of four diseases, the model's average performance exceeds that of all others, while is second for the fourth disease. We next present a discussion on each disease class individually for the model.

|  |  | NIH Baseline | CheXpert-NET | Shifa-NET | CheXpert-Shifa-NET |
|---|---|---|---|---|---|
| **Atelectasis** | **NIH Test** | 0.767 | **0.808** | 0.781 | 0.791 |
|  | **Shifa Test** | 0.746 | 0.741 | 0.779 | **0.781** |
|  | **CheXpert Val** | 0.800 | 0.826 | 0.790 | **0.833** |
|  | **Average** | 0.771 ± 0.027 | 0.792 ± 0.045 | 0.783 ± 0.006 | **0.801 ± 0.026** |
| **Cardiomegaly** | **NIH Test** | 0.847 | 0.851 | **0.867** | 0.860 |
|  | **Shifa Test** | 0.826 | 0.852 | 0.903 | **0.903** |
|  | **CheXpert Val** | 0.813 | 0.797 | 0.806 | **0.815** |
|  | **Average** | 0.827 ± 0.017 | 0.833 ± 0.031 | 0.859 ± 0.049 | **0.859 ± 0.044** |
| **Pleural Effusion** | **NIH Test** | 0.848 | **0.861** | 0.848 | 0.850 |
|  | **Shifa Test** | 0.844 | 0.852 | **0.882** | 0.880 |
|  | **CheXpert Val** | 0.857 | **0.926** | 0.906 | 0.919 |
|  | **Average** | 0.850 ± 0.007 | 0.880 ± 0.040 | 0.879 ± 0.029 | **0.880 ± 0.049** |
| **Pneumonia** | **NIH Test** | 0.707 | **0.762** | 0.723 | 0.724 |
|  | **Shifa Test** | 0.696 | 0.716 | 0.795 | **0.797** |
|  | **CheXpert Val** | 0.791 | 0.748 | **0.796** | 0.768 |
|  | **Average** | 0.731 ± 0.051 | 0.742 ± 0.023 | **0.771 ± 0.042** | 0.763 ± 0.037 |

*Table 7: Consolidated results*

## 4.1 Atelectasis

In the case of Atelectasis, we observe on average that the CheXpert-Shifa-NET model shows a marked improvement. It shows a strong performance on the Shifa test set, and comparably better performance on the CheXpert validation set then CheXpert-NET. In the case of the NIH testing set, the CheXpert-NET model outperforms all the other experimental models, with CheXpert-Shifa-NET being a close second, however, the performance of CheXpert-NET shows high variability over the various test sets, while CheXpert-Shifa-NET is consistently ranked either as best or close to best. On average, CheXpert-Shifa-NET model outperforms all other models.

## 4.2 Cardiomegaly

In the case of Cardiomegaly, we observe that all the models are relatively consistent and on average the CheXpert-Shifa-NET model demonstrates the top performance in comparison to all the models. The largest performance improvement is observed in Shifa test datasets. Moreover, the Shifa-NET model performs well for predicting Cardiomegaly, potentially indicating that CheXpert-Shifa-NET gains its performance on Cardiomegaly from being trained on Shifa training dataset. Interestingly, the CheXpert-NET model performs worse on CheXpert dataset, also indicating that training on CheXpert dataset is less valuable for Cardiomegaly predictions.

## 4.3 Pleural Effusion

In the case of Pleural Effusion, we observe that all three models perform comparably well on average outperforming the NIH baseline model. The performance improvements are significant on both the Shifa test set and the CheXpert validation set. The scores of all four models on the NIH testing set are fairly similar, with the CheXpert-NET model slightly outranking the other three. On the CheXpert set, all three models exhibit a significant improvement as compared to the baseline model, with the within dataset performance of the CheXpert-NET model being strongest, and the cross-performance of the CheXpert-Shifa-NET dataset coming a close second.

## 4.4 Pneumonia

In the case of Pneumonia, Shifa-NET model exhibits a strong performance on all three testing sets with the average performance of CheXpert-Shifa-NET model being close to Shifa-NET. The CheXpert-NET model exhibits strong performance only on the NIH testing set, and once again shows substantial variation across the test sets. However, its important to note that the CheXpert validation set contains only nine pneumonia positive cases, which is too few to draw conclusive results.

Our results show that all three models CheXpert-Shifa-NET model, Shifa-NET model and CheXpert-NET perform significantly better than the NIH baseline (t-test[29] p-value 0.00441, 0.00785 and 0.0414, Wilcoxon test[30] p-value 0.00379, 0.00639 and 0.04182), confirming that training DenseNet-121 model on additional

datasets improves the prediction performance.

## 4.5 Visualization and model interpretability

In this section, we qualitatively evaluate the predictions of CheXpert-Shifa-NET and demonstrate cases where our model accurately predicts and localize the disease classes. To evaluate the models interpretability and capability to aid clinical decision making, we identify and plot the activation map as a representation of the models view of features relevant to prediction. Specifically, we use the gradient class activation map (GradCAM)[31] to plot the relevant X-Ray regions. GradCAM uses the gradients flowing into the last convolutional layer to produce a localization map. Each channel in the final convolutional feature map is weighted with the gradient of the class with respect to the channel. This allows us to visualize the regions the model finds relevant for computing predictions for each of the individual classes. Figure 3 shows randomly sampled examples predicted by our model and the correspondingly identified regions. We obtain expert evaluation of the model predictions by radiologist.

In Figure 3A, our model correctly classifies this as a pneumonia positive case. The colormaps depict an overcalling of the pathology but the major findings are consistent with the findings of the radiologist. In Figure 3B, the model correctly identifies and localizes around the enlarged heart (cardiomegaly) very accurately. In Figure 3C, while our model is correctly classifying both pathologies, the colormaps for the pleural effusion class are not where they would be expected, however, for the pneumonia class the model not only classifies the chest x-ray correctly, it is also identifying the areas of concern very accurately. We note that while our model was able to correctly classify pleural effusion very accurately, it had difficulty in localizing it. Lastly, in Figure 3D, the model correctly classifies and localizes both cardiomegaly and pneumonia. The strong predictive generalizability of the model coupled with precise identification of the disease regions demonstrates the opportunity for potential clinical applicability of generalized deep learning in radiological diagnosis using chest X-Rays.

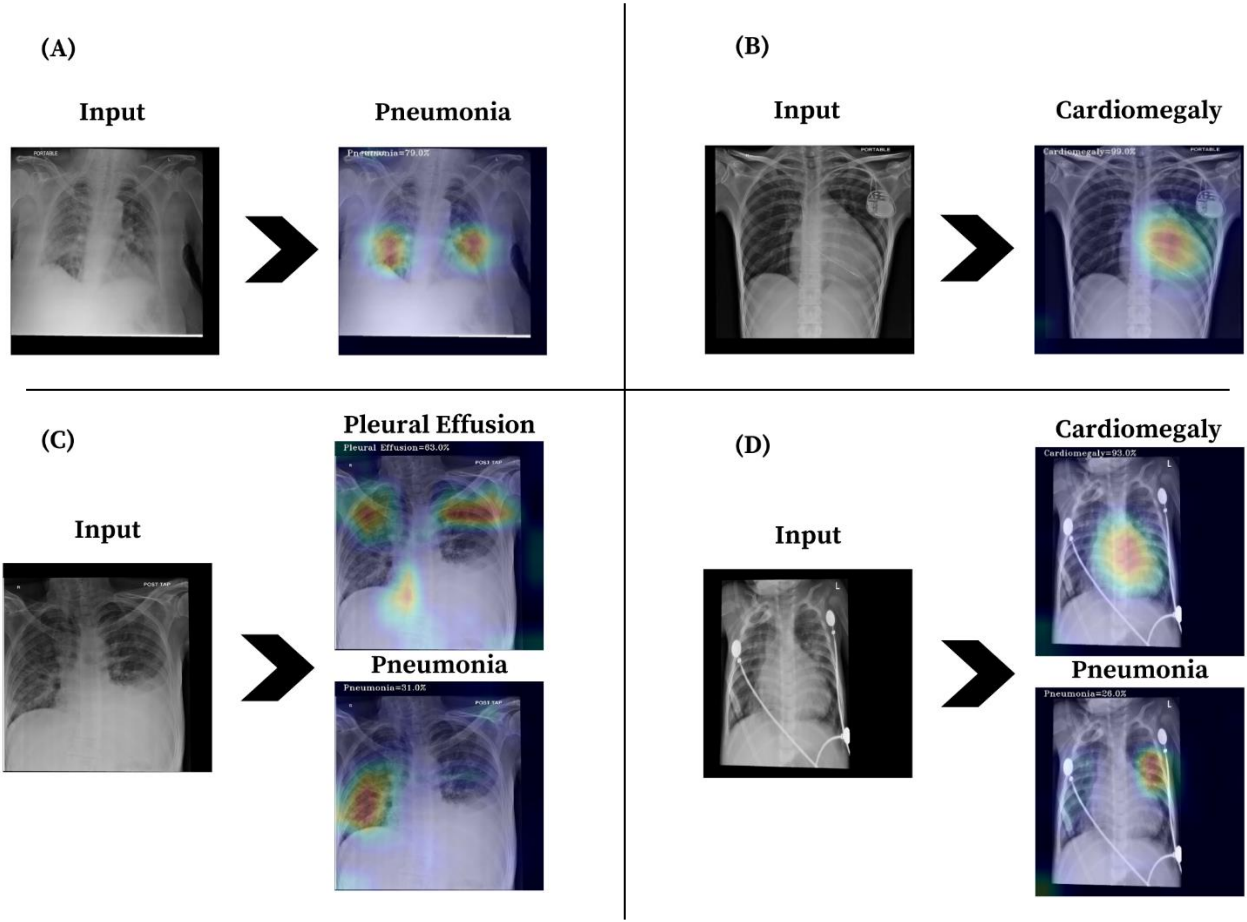

*Figure 3: Grad-CAM Results*

## 5 Discussion

Our primary goal in this work was to investigate the generalization performance of deep learning models to datasets from multiple institutions and populations. Our experimental results signify that deep learning models trained in multiple stages across data from multiple health institutions can generalize for multiple disease classes. The sequential nature of the training makes this approach suitable for continual improvements without the need to retrain from scratch. For 3 out of the 4 disease classes, the final generalized model, CheXpert-Shifa-NET, trained on all three datasets, shows the highest performance on average across the testing sets from all three healthcare institutions. These results are important because they demonstrate both, generalization ability, as well as precise identification of localized regions relevant for prediction, opening the opportunity for automatic radiological screening in a clinical settings. Interestingly, for Atelectasis and Cardiomegaly disease classes, our model shows an improvement in performance when trained according to our proposed methodology. It is important to note that one of the three datasets used in our work was from an entirely different region and population, and it is hard to account for confounding variables in that scenario. However, we note that training on this dataset does not degrade the performance of the model on any of the other datasets, rather, it leads to increased performance in a majority of cases, attesting the generalizability of the model. Another important finding which supports our claim of generalization across multiple health institutions is the performance of our generalized model on the CheXpert test set. This test set consists of a radiologist annotated ground truth and the training on SIH dataset in the did not degrade the generalization performance on CheXpert dataset indicates the generalization potential of the CheXpert-Shifa-NET model. We assess that this level of generalization across multiple disease classes is in line and goes a step further from the recent findings within similar population types[22,23].

# 6    Conclusion

We conclude that deep learning models can generalize for the case of multiple disease classification from chest x-rays, although the extent of the generalization performance is limited and varies for each disease class. We note that the approximate nature of labeling the radiology reports through automated labeling techniques may impact the performance of the classifier for some disease classes, however, the methodology used in this line of work reduces the manual labeling effort, while still maintaining reasonable accuracy levels on test datasets across the multiples testing sites and disease classes. For the case of clinical diagnostics, this approach does require evaluation by certified radiologists; however, visualizing the identified regions of the generalized chest x-ray image classification model can quickly provide a radiologist with an accurate depiction of a disease class, potentially make diagnosis faster.